**Atomistic insights into solid solution strengthening: size misfit versus stiffness misfit**

Aoyan Liang[a,b], Nicolas Bertin[a], Xinran Zhou[a,c,d], Sylvie Aubry[a], Vasily V. Bulatov[a*]

[a] *Lawrence Livermore National laboratory, 7000 East Avenue, Livermore, CA, USA*
[b] *University of Southern California, 3551 Trousdale Parkway, Los Angeles, CA, USA*
[c] *University of California Los Angeles*, *405 Hilgard Avenue, Los Angeles, CA, USA*
[d] *Argonne National Laboratory, 9700 S Cass Ave, Lemont, IL, USA*
*Correspondence: bulatov1@llnl.gov

**Summary**

Used for centuries to enhance mechanical properties of materials, solid solution strengthening (SSS) is a classical metallurgical method in which small amounts of impurity elements (solutes) are added to a base metal (solvent). Developed for dilute alloys, classical theories of SSS are presently challenged by the ongoing explosive development of complex concentrated alloys (CCA) in which all component elements are present in nearly equal fractions. Here we develop a method of "computational alchemy" in which interatomic interactions are modified to continuously and systematically vary two key parameters defining SSS – atomic size misfit and elastic stiffness misfit – over a maximally wide range of misfit values. The resulting model alloys are subjected to massive (~ $10^8$ atoms) Molecular Dynamics (MD) simulations reproducing full complexity of plastic strength response in concentrated single-phase body-centered cubic (BCC) solid solutions. At variance with views prevailing in the literature, our computational experiments show that stiffness misfit can contribute to SSS on par if not more than size misfit. Furthermore, depending on exactly how they are combined, the two misfits can result in synergistic (amplification) or antagonistic (compensation) effect on alloy strengthening. Taking advantage of full atomistic resolution of our simulations, we use *in silico* computational microscopy to examine which mechanisms control plastic strength of our model alloys. Our data suggests that, just like in unalloyed BCC metals, motion of screw dislocations plays a dominant role. In contrast to real CCAs in which every constituent element comes with its specific combination of atomic size and elastic stiffness, our alchemical model alloys sample the space of misfit parameters continuously thus augmenting the much more constrained and inevitably spotty experimental exploration of the CCA design space. Taking advantage of unique to our approach ability to define alloy misfit parameters, our computational study demonstrates how useful insights can be gained from intentionally unrealistic alchemical models. Rather than practical recommendation for alloy design, our computational experiments should be regarded as a proving ground for further SSS theory development.

**Introduction**

Solid solution strengthening (SSS) is an essential metallurgical method in which solute atoms are introduced into a host metal to induce local distortions in its crystal lattice thereby impeding dislocation motion and strengthening the material. Classic theory [1,2] and experiments [3–7] attribute SSS to two key factors: (1) size misfit referring to the difference in atomic radii and (2) stiffness misfit referring to the difference in elastic moduli between solute elements and the host metal. Over the last two decades, complex concentrated alloys (CCAs) have emerged as promising materials in large part due to their demonstrated exceptional mechanical properties [8–10]. CCAs typically consist of four, five or even more elements mixed together in large fractions so that no single element can be considered host [11,12]. Given the vast compositional space of CCAs, understanding physical mechanisms of CCA strengthening is essential for developing materials with superior mechanical properties. A number of theoretical models of CCA strength have been developed generally along the lines of the classic SSS theory originally developed for dilute alloys [1,2,13–19]. Toda-Caraballo and Rivera-Diaz-del-Castillo [17] adapted Labusch's statistical theory for binary systems to predict the yield strength of CCAs by refining the formulation of size and stiffness misfits. Varvenne et al. [13] and Maresca et al. [19] proposed models in which each atom in the CCA is regarded as a solute embedded in an effective matrix of an "averaged host material", with strengthening arising from dislocation interactions with local concentration fluctuations around the average composition. In addition to physics-based models, Wen et al. [20] employed machine learning to extract factors contributing to SSS. However, despite significant recent progress in theory and simulations, mechanisms of strengthening in complex alloys are still not fully understood. In particular, even if size and stiffness misfits are widely recognized as two essential factors defining CCA strength, their relative importance for and individual and combined contributions to strengthening remain uncertain.

It is difficult if not impossible to decouple effects of size and stiffness misfit in experiments where every constituent element in a CCA comes with its own fixed combination of two misfit parameters that cannot be independently varied (see Fig. 1 for a plot of size-stiffness combinations among common alloying elements). Here we propose a computational approach for gaining physical insights into SSS in which atomic size and elastic stiffness of alloying elements can be varied continuously, either individually or in any desired combination. Starting from a base interatomic potential (IP) describing atom-atom interactions in tantalum [21], we modify the base IP to create "virtual" elements with arbitrary combinations of atomic size and elastic stiffness (see Methods). The so-created virtual elements are then combined in a desired proportion to form an "alchemical" model CCA subsequently subjected to deformation in a large-scale (~55 million atoms) Molecular Dynamics (MD) simulation. Over one hundred such large-scale MD simulations are performed to observe and quantify strengthening as a function of size and stiffness misfits. In contrast to real CCAs in which every constituent element comes with its specific combination of atomic size and elastic stiffness (Fig. 1), our alchemical model alloys cover the 2D space of misfit parameters continuously thus augmenting the much more constrained and inevitably spotty and sparse experimental exploration of the CCA design space. By creating virtual alloys with controlled misfits, we aim to clarify relative importance of size and stiffness misfits in SSS as well as to quantify their combined effects. To the extent that our MD computational experiments are representative of strength response of real CCAs, insights gained from our alchemical exploration provide a deeper quantitative understanding of fundamental mechanisms of alloy strengthening and offer valuable benchmark data for further SSS theory development.

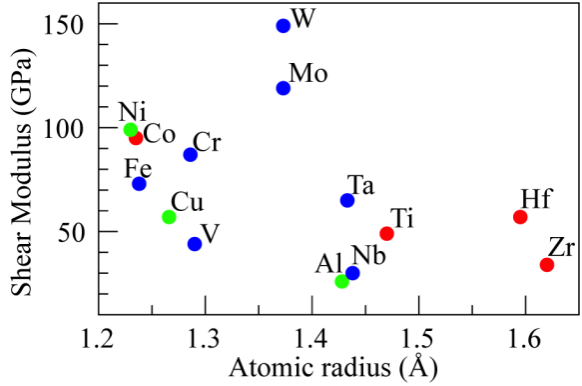

**Figure 1. Relationship between atomic radius and shear modulus among common alloying elements.** Each element is color-coded according to its stable crystal structure at room temperature (Blue: Body-centered cubic (BCC), Green: Face-centered cubic (FCC), Red: Hexagonal close-packed (HCP)). Data obtained from the Materials Project [22].

**Results and Discussion**

In order to reveal effects of size and stiffness misfits most directly, in creating alchemical elements for our model alloys (see Methods) we placed several constraints on continuous modifications of the base interatomic potential. In addition to an obvious condition that atomic fractions of elements comprising an alloy must sum up to 1.0, we create and combine our alchemical elements so as to maintain atom size and elastic stiffness averaged over alloy elements exactly the same as in the base metal, i.e. Embedded Atom Model (EAM) model of Ta. Here average atom size $r_a$ and average stiffness $\mu_a$ are defined as

$$r_a = \sum_{i=1}^{n} c_i r_i, \qquad \mu_a = \sum_{i=1}^{n} c_i \mu_i \qquad (1)$$

whereas size misfit $\Delta r$ and stiffness misfit $\Delta \mu$ are defined as [20]

$$\Delta r = \sqrt{\sum_{i=1}^{n} c_i \left(1 - \frac{r_i}{r_a}\right)^2}, \qquad \Delta \mu = \sqrt{\sum_{i=1}^{n} c_i \left(1 - \frac{\mu_i}{\mu_a}\right)^2} \qquad (2)$$

where $c_i$, $r_i$, and $\mu_i$ are the atomic fraction, atom size, and stiffness of element $i$, respectively. Observing that the base metal Ta corresponds to $\Delta r = \Delta \mu = 0$, any difference in strength from the base metal predicted for a model alloy created under such constraints is attributed directly to the misfits.

Our first series of MD simulations was of binary equiatomic model alloys in which size misfit was kept at zero, $\Delta r = 0$, but stiffness misfit $\Delta \mu$ was raised from one alloy to the next in increments of 0.05. Stress-strain response curves recorded in MD simulations of model alloys are shown Fig. 2a and strength values extracted from the same stress-strain data are plotted as brown circles in Fig. 2b. Strength values reported here and everywhere in the following were computed by averaging flow stress over the interval of true strains from 0.15 to 0.40. Observing that alloy strength rises monotonically with increasing stiffness misfit, it is of interest to see if stiffness misfit parameter $\Delta \mu$, as defined in eq. (2), is a reliable predictor of the amount of strengthening. With this in mind, we combined some of the same alchemical elements in non-equiatomic binary and ternary model alloys and simulated their strength response in MD. All in all, 34 model alloys with stiffness misfit only were simulated and their predicted strength values are shown in Fig. 2b (several model alloys had similar or even identical values of stiffness misfit and yet different number of components and/or compositions). Plastic strength in MD simulations is observed to fall close to a single line confirming $\Delta \mu$ as an accurate predictor of alloy strengthening regardless of the number of elements and element fractions in a model alloy.

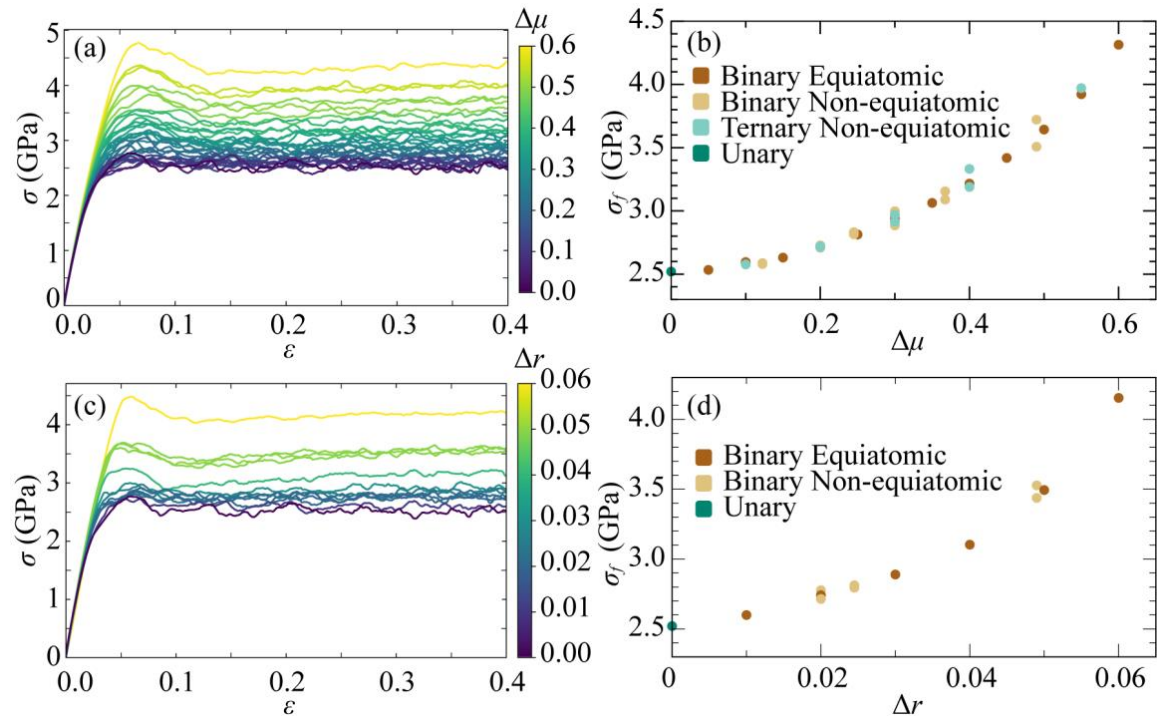

**Figure 2. Effects of solo stiffness and size misfits.** Stress ($\sigma$) – strain ($\varepsilon$) response of alchemical alloys with (a) only stiffness misfits, $\Delta \mu$, and (c) only size misfits, $\Delta r$, subjcted to compression in

MD simulations. The curves are color-coded according to misfit values in each simulated model alloy, with color maps at the right of each plots. Flow stress, $\sigma_f$, as a function of (b) stiffness misfit and (d) size misfit.

The largest stiffness misfit examined in this series of MD simulations was 0.6 because model alloys with any greater stiffness misfit were observed to amorphize under uniaxial compression. Previous studies have attributed spontaneous amorphization to size misfit [23–25], however our simulations reveal that, even in the absence of atom size differences, stiffness misfit alone can destabilize a crystal and trigger amorphization under deformation.

We performed a similar series of MD simulations of binary equiatomic and non-equiatomic model alloys in which two component elements had the same stiffness ($\Delta\mu = 0$) but different atom sizes ($\Delta r \geq 0$). The results are presented in Figs. 2c and 2d. Similar to stiffness misfit, size misfit strengthens the alloys and the amount of strengthening is accurately predicted by the size misfit parameter both for equiatomic and non-equiatomic alloys. However, the range of permissible size misfits over which our model binary alloys remain stable in the BCC structure is approximately ten times narrower than that for the stiffness misfit. Binary alloys with size misfit exceeding 0.06, are observed to amorphize under uniaxial compression.

Within their corresponding ranges of stability against amorphization, contribution of two misfits to strengthening is comparable with stiffness misfit producing slightly more strengthening. As a point of comparison, maximum size misfit achievable in an alloy comprised of refractory metals Nb, Mo, Ta, W and Re, is $\Delta r_{max} = 0.032$, whereas maximum stiffness misfit achievable within the same alloy family is $\Delta\mu_{max} = 0.65$. This size misfit is close to about half of the maximum permissible size misfit (before amorphization kicks in) while the latter stiffness misfit slightly exceeds the maximum permissible stiffness misfit of 0.60. Data shown Figs. 2b and 2d suggests that maximum relative strengthening achievable within the refractory family of alloys solely due to size misfit is by a factor 1.2, whereas maximum achievable strengthening solely due to stiffness misfit is at least by a factor 1.72. This is at substantial variance with the literature where most theoretical models of SSS generally ascribe greater strengthening effect to size misfit, with stiffness misfit completely neglected in some of the more recent models [7,13,18,26,27].

Given that most elemental metals differ both in atom size and in stiffness, combined effect on strengthening of the two misfits has been substantially covered in the literature focused on dilute alloys [1,2,28–31]. Here, rather than appealing to any theoretical considerations, we use our method to generate model alchemical elements of variable atom sizes and elastic stiffnesses so as to systematically assess and quantify combined effect of two misfits on alloy strength predicted in large-scale MD simulations. When it comes to combining two misfits, there are several, generally different ways in which elemental metals can be combined in an alloy. Considering an alloy comprised of $n$-elements mixed together at fractions $c_i$ ($i = 1, 2, \ldots, n$), each element is either smaller or larger than the average atom size defined in eq. (1) and either softer or stiffer than the average shear modulus defined in the same equation. Thus, with respect to the two averages $r_a$ and

$\mu_a$, each constituent element can have one of four distinct types of the misfit pair: small-soft, small-stiff, large-soft, large-stiff (see Fig. 3). Obviously only two misfit pair types can be represented in a binary alloy for a total of two possible distinct pair type combinations labelled B1 and B2 in the figure. Three misfit pair types can be represented in a ternary alloy adding four more distinct type combinations labelled T1, T2, T3 and T4. In alloys comprised of $n \geq 4$ elements all four misfit pair types can be simultaneously represented adding one more combination Q for a total of seven distinct pair type combinations. To limit combinatorial complexity rising with further increase of the number of components $n$, all component elements of the same misfit pair type can be lumped into a "consolidated component element" with consolidated fraction $\tilde{c}$, consolidated atom radius $\tilde{r}$ and consolidated shear modulus $\tilde{\mu}$ defined as

$$\tilde{c} = \sum_{i=1}^{k} c_i, \qquad \tilde{r} = \sum_{i=1}^{k} \frac{c_i r_i}{\tilde{c}}, \qquad \tilde{\mu} = \sum_{i=1}^{k} \frac{c_i \mu_i}{\tilde{c}} \qquad (3)$$

where the sums are taken over all component elements of the same misfit pair type. At the expense of slight undercounting of size and stiffness misfits while preserving average atom size and average stiffness, lumping alloy components in this manner limits the number of consolidated components to no more than four.

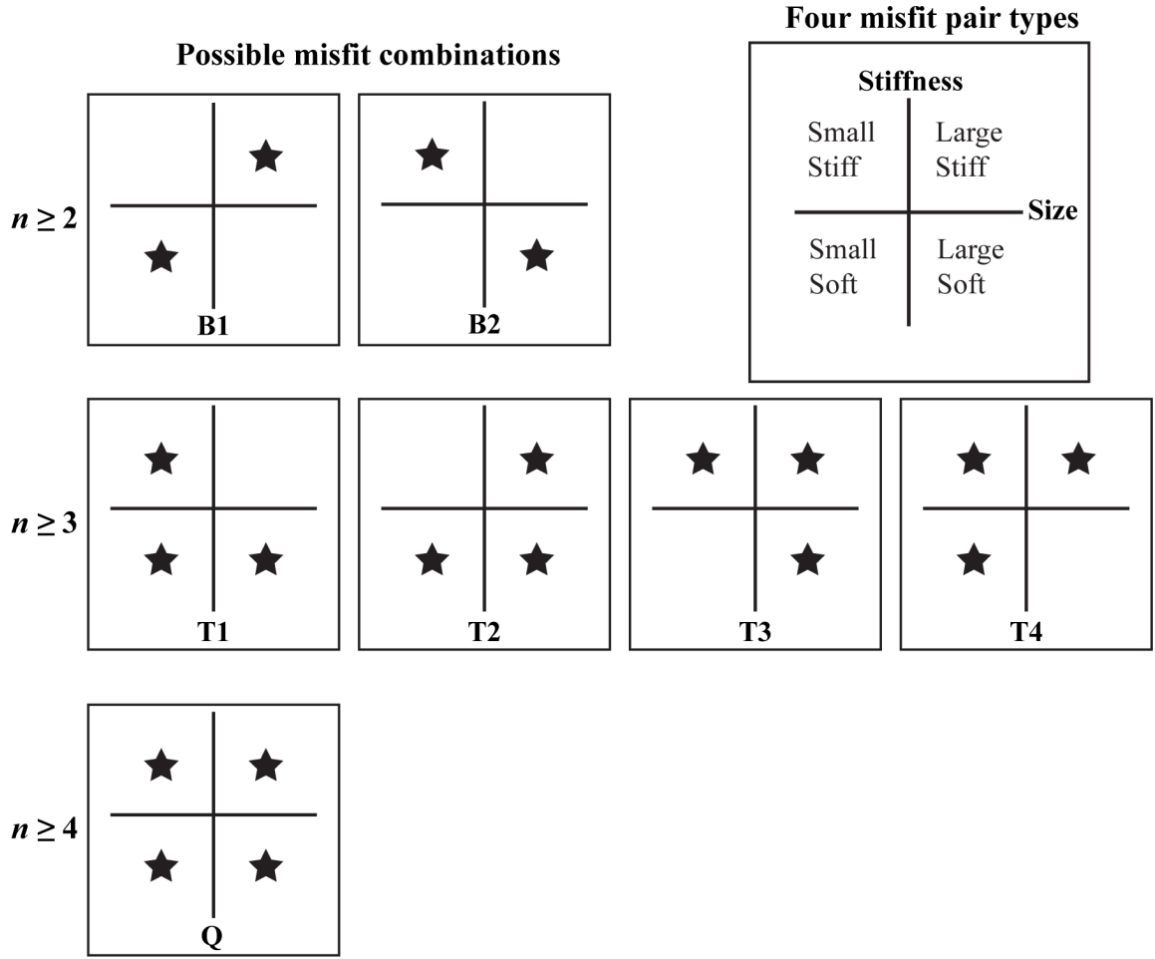

**Figure 3. Schematic of seven distinct misfit pair type combinations.** The stars placed in four quadrants represent one of four distinct misfit pair types defined in the diagram on the upper right: small-soft, small-stiff, large-soft, and large-stiff. The origin (0,0) corresponds to atom size and elastic stiffness averaged over all alloy components. Considering misfit pair types of their consolidated components, all multi-component alloys can be assigned to one of seven distinct types: two binary types B1 and B2, four distinct ternary types T1, T2, T3 and T4 or the quaternary type Q.

Even with extraordinary computational resources allocated to us, we could not afford a large number of large-scale MD simulations required to explore the entire multi-dimensional space of all misfit pair types. Here we opt to systematically examine only equiatomic binary model alloys for which variations in predicted strength can be conveniently shown on a 2D plane with two coordinates corresponding to two misfits. All in all, including all alloys with only one "solo" misfit described earlier in Fig. 2, we created and simulated 91 binary model alloys to systematically sample strength response over the misfit plane in steps of 0.01 from 0 to 0.06 along the size misfit axis and in steps 0.1 from 0 to 0.6 along the stiffness misfit axis. Several B1 alloys combining large size and large stiffness misfits amorphized under deformation, thus further narowing the range of misfit space in which alloy strength response could be simulated (also see Fig. 5a). No such instability was observed among B2 alloys within the entire explored range of two misfits.

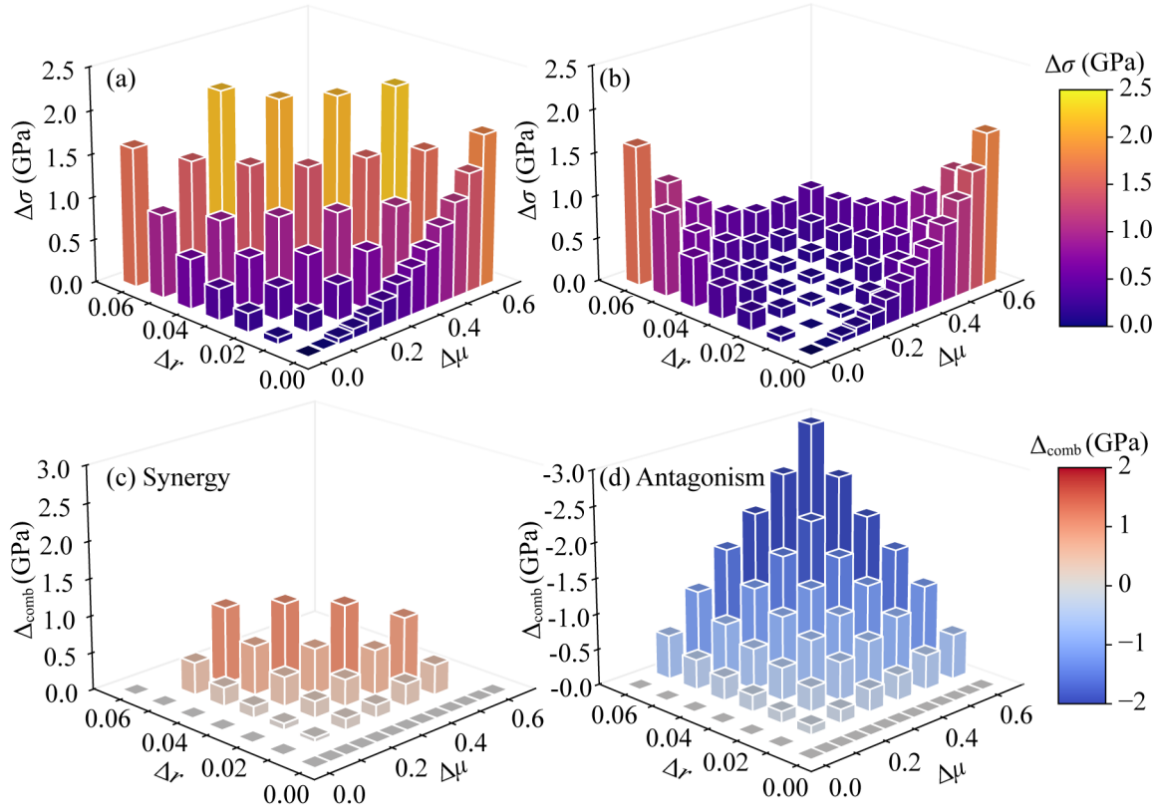

**Figure 4. Effect of combined size misfit $\Delta r$ and stiffness misfit $\Delta \mu$ on alloy strengthening.** Heights and colors of bars show how much the flow stress ($\Delta\sigma$) increases in B1 alloys (a) and B2 alloys (b) over the flow stress of the reference unalloyed Ta corresponding to the zero strengthening baseline. Excess strengthening ($\Delta_{comb}$) in B1 alloys (c) and B2 alloys (d) defined as the difference between strengthening induced by two misfits combined and the sum of their solo strengthening effects previously shown in Fig. 2. Note the negative scale of excess strengthening in (d).

Calculated as $\Delta\sigma = \sigma_{alloy} - \sigma_{Ta}$, where $\sigma_{alloy}$ and $\sigma_{Ta}$ represent flow stress of the alloy and the reference elemental Ta, respectively, predicted alloy strengthening over the 2D misfit plane is shown in Figs. 4a and 4b. To observe the effect of various misfit pairs still more clearly, Figs. 4c and 4d show simulated strengthening in excess of additive strength computed by summing strengthening induced by each of the two solo misfits (from data shown in Fig. 2). Consistent with Fig. 2, size and stiffness misfits contribute significantly to strengthening in both B1 and B2 quadrants, however the magnitude of SSS in the two quadrants is strikingly different. In quadrant B1, combined effect of size and stiffness misfits is clearly synergistic (amplified), meaning that net strengthening exceeds the sum of two solo contributions presented in Fig. 2. In contrast, two misfits are antagonistic (compensated) in quadrant B2 where combined strengthening is markedly smaller than the sum of two solo effects. Furthermore, alloys with greater stiffness and size misfits exhibit greater synergism and greater antagonism in quadrant B1 and quadrant B2, respectively. These observations indicate that greater SSS can be achieved by combining small-soft metals with large-stiff metals, rather than in the reverse combination.

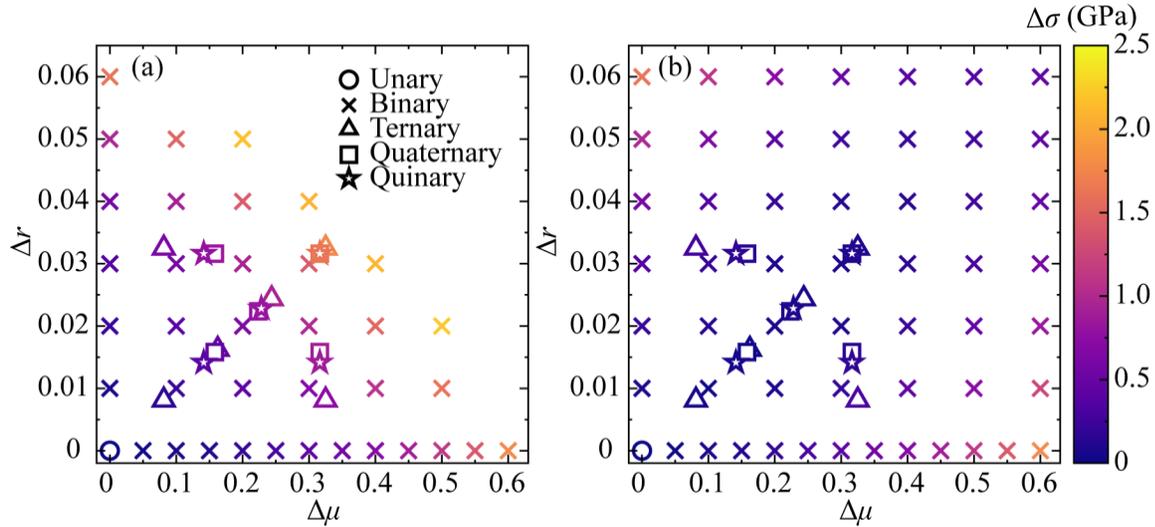

**Figure 5. Strengthening Δ$\sigma$ in equiatomic binary and multicomponent alloys** of misfit pair types falling into quadrant B1 (a) and quadrant B2 (b). Symbols are color-coded according to Δ$\sigma$ values. Note an outer boundary on the misfit plane in quadrant B1 beyond which all model alloys are found to be unstable against amorphization.

To examine if and how much specific combinations of two misfits are predictive of strengthening, we added several MD simulations of equiatomic multi-element model alloys with up to five element species in which values of atom size and elastic stiffness were assigned among the components so that the resulting alloys could still be attributed to either quadrant B1 or quadrant B2. According to the attribution rules defined in Fig. 3, a multicomponent alloy is attributed to quadrant B1 if its every element is either smaller-softer or larger-stiffer than the fraction-averaged atom size and stiffness. Conversely, a multicomponent alloy is attributed to quadrant B2 if its every element has a misfit pair of either smaller-stiffer or larger-softer. As shown in Figs. 5a and 5b, strength predicted for multi-component alloys falling into two quadrants is close to strength predicted for binary alloys with the same combinations of two misfits suggesting that, by itself, the number of elements in an alloy does not significantly affect SSS. The key implication is that comparable SSS should be achievable in binary or multi-component alloys of similar integral misfits regardless of the number of components.

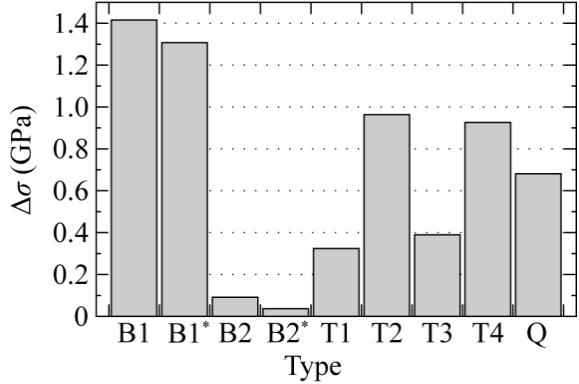

**Figure 6 Strengthening Δσ in seven distinct alloy types with comparable misfit values ($\Delta r$, $\Delta \mu$).** B1, B2 and Q have misfit values of (0.03, 0.3), while T1-T4 have slightly lower misfit values of (0.02814, 0.2814). B1$^*$ and B2$^*$ represent interpolated binary alloy data from Fig. 4 with the same reduced misfit as T1–T4.

Even after reducing effective dimensionality of the space of alloy misfit pairs to four consolidated components, it would have taken of the order of $10^4$ large-scale MD simulations to explore the entire 4D space of misfit pair types to the same degree of detail as we did for our equiatomic model binary alloys (Fig. 4). Here we limit ourselves to just one representative equiatomic alloy from each of the seven alloy types depicted in the diagram in Fig. 3 for a total of five additional MD simulations. Simulated strengthening of two binary alloys of the B1 and B2 types, four ternary alloys of the T1, T2, T3 and T4 types and a quaternary alloy of the Q type are compared in Fig. 6. In assembling an equiatomic quaternary alloy of the Q type, we re-used four alchemical EAM potentials previously generated to examine SSS in equiatomic binary alloys with misfits $\Delta r = 0.03$ and $\Delta \mu = 0.3$ thus keeping both misfits in the Q alloys the same. Likewise, we generated equiatomic ternary alloys of T1, T2, T3 and T4 types using previously generated alchemical elements. However, to keep average atom size and average stiffness of four ternary alloys unchanged we had to allow a small reduction in the overall size misfit from 0.03 to 0.02814 and in the stiffness misfit from 0.3 to 0.2814. Given smooth and systematic variations of strength with two misfits shown in Fig. 4, such small differences in the overall misfits can be safely ignored in comparing strengthening induced in alloys of various types. Simulated strength of alloys of five additional types depicted in Fig. 3 is compared to strength of two binary alloys with the same misfits in Fig. 6. This comparison adds support to our earlier observation that combining small-soft and large-stiff components increases strength much more than combining large-soft and small-stiff components. Furthermore, adding components of the latter two misfit pair types is not beneficial for strengthening of any multi-component alloy, regardless of the number of alloy components.

SSS has been previously attributed to deviations of atom positions from a perfect lattice caused by local variations in chemical composition and impeding dislocation motion [32–37]. To see if strengthening predicted in our MD simulations is covariant with lattice distortions induced by local composition fluctuations, we computed the root mean square deviation (RMSD) of atomic coordinates from a perfect BCC lattice positions in model alloys comprised of 2000 atoms distributed over the BCC lattice sites at random in equal fractions. After minimizing the crystal's energy at 0K, the RMSD was computed for each alloy as

$$\text{RMSD} = \sqrt{\frac{1}{3N}\sum_{i=1}^{3N}(X_i - X_i^0)^2} \qquad (4)$$

where $\boldsymbol{X}$ and $\boldsymbol{X^0}$ are the atomic coordinates in an alloy distorted due to size and stiffness misfits and coordinates of the same atoms in a perfect crystal with the same lattice constant, respectively.

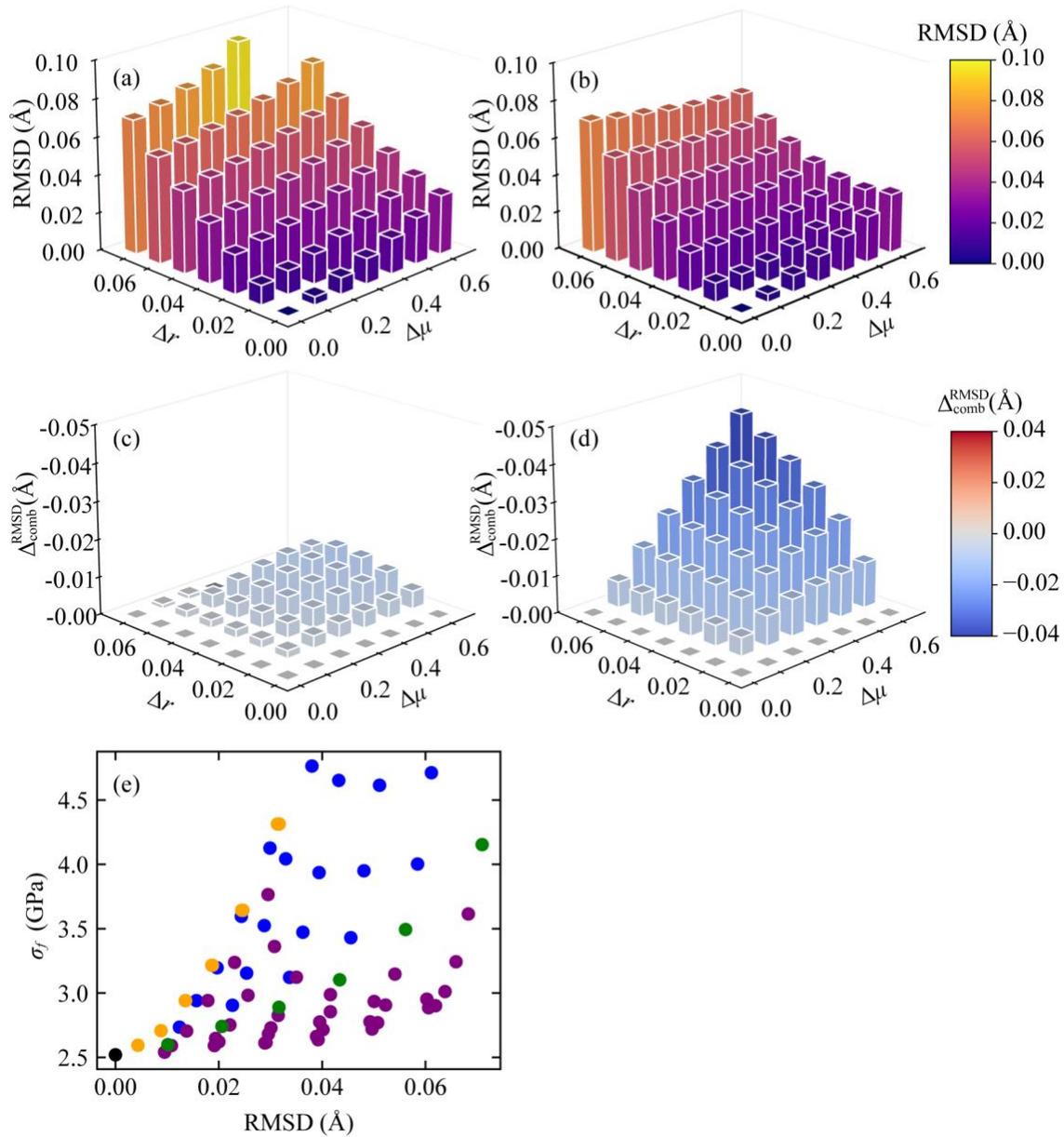

**Figure 7. Effect of combined size misfit $\Delta r$ and stiffness misfit $\Delta \mu$ on RMSD.** RMSD of atomic positions from a perfect, undistorted crystal as a function of size and stiffness misfits for alloy compositions of type B1 (a) and type B2 (b). Excess RMSD in B1 alloys (c) and B2 alloys (d) defined as the difference between RMSD induced by two misfits combined and the sum of RMSD induced separately by each of two solo misfits. Note the negative scales of excess RMSD in (c) and (d). (e) Flow stress ($\sigma_f$) against RMSD computed for all alloys compositions presented in Fig. 4. Symbols are colored according to misfit types: no misfit – black, misfit of B1 type – blue, misfit of B2 type - purple, solo stiffness misfit - orange, solo size misfit - green.

Figure 7 shows the RMSD as a function of size and stiffness misfits computed for the same 91 equiatomic binary alloys for which strengthening was computed and plotted in Fig. 4. Similar to strengthening, two different ways to combine the misfits induce markedly different variations in RMSD in quadrants B1 and B2. In quadrant B1 the RMSD rises monotonically with increasing misfits (Fig. 7a), however combined effect of two misfits on RMSD is not synergistic as in strength, but moderately antagonistic (Fig. 7c). In quadrant B2 the RMSD varies non-monotonically with increasing size misfit while combined effect of two misfits on the RMSD is clearly antagonistic. Comparison of Fig. 4 and Fig. 7 suggests that the RMSD is not an accurate predictor of strengthening in our model alloys. This conclusion is further supported by Fig. 7e showing little if any covariance between predicted alloy strength and its RMSD. For example, model alloys with nearly the same RMSD ~ 0.04 Å exhibit widely varied strengthening from almost no strengthening for some alloys to maximum strengthening for other alloys.

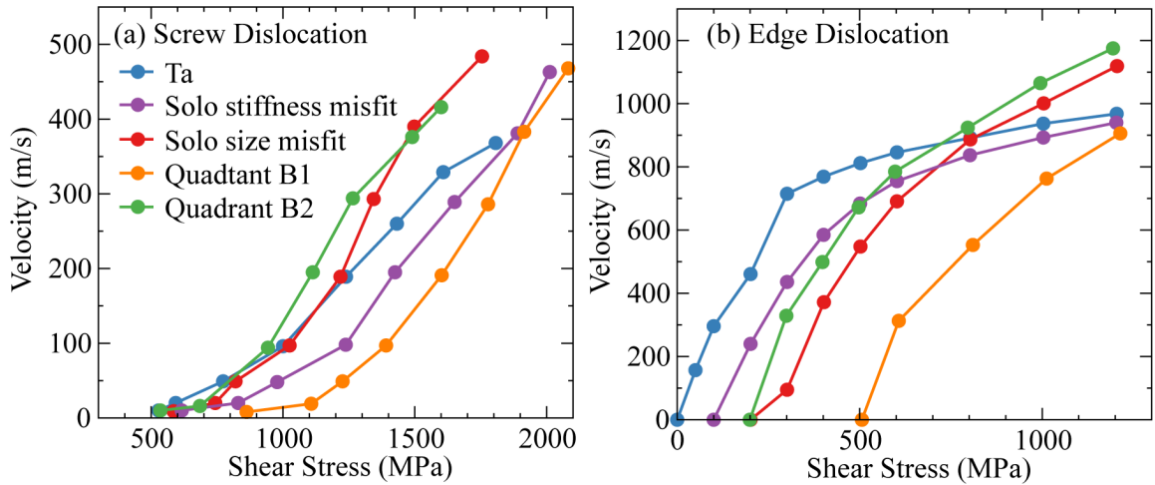

**Figure 8. Dislocation mobility in different alloy types.** (a) Screw and (b) edge dislocation mobility calculated for the elemental reference Ta model (blue) and four representative model alloys: one alloy of the B1 type (orange) and one alloy of the B2 type (green) both with size and stiffness misfits $\Delta r = 0.03$ and $\Delta\mu = 0.3$ and two alloys with solo misfits - solo size misfit $\Delta r = 0.03$ (red) and solo stiffness misfit $\Delta\mu = 0.3$ (purple).

Crystal plasticity response results form, but is not reducible to dislocation motion and entails other important mechanisms, e.g. dislocation multiplication and dislocation interactions. However, contribution to strength of various collective mechanisms of dislocation behavior is difficult to assess experimentally and theoretically, thus nearly all existing theoretical models and computer simulations of SSS assume, explicitly or implicitly, that strengthening results solely from the interaction of individual dislocations with a sea of alloying solutes more or less randomly distributed over the lattice sites. No such assumption is made in our massive MD simulations in which mobility of individual dislocations and collective mechanisms of dislocation interactions are all naturally accounted for. Thus, it is of interest to assess if and how much mobility of individual dislocations contributes to alloy strength. With this purpose in mind we performed MD simulations of individual screw and edge dislocations in five model materials: the reference elemental Ta and four equiatomic binary alloys: an alloy with solo size misfit $\Delta r = 0.03$, an alloy with solo stiffness misfit $\Delta \mu = 0.3$ and two alloys with the same misfits (0.03, 0.3) combined in B1 and B2 alloy types. Comparing Figs. 4 and 8, we generally observe close qualitative correspondence between effects of alloying on plastic strength and on dislocation mobilities. In particular, the synergistic combination of two misfits reduces mobilities of both edge and screw dislocations much more than does the antagonistic combination of the same misfits. Also worth noting that, although both misfits – taken solo or in combination – reduce mobilities of both dislocation types, size misfit has a much greater effect on the edge dislocations whereas stiffness misfit has a greater effect on the screws. Notably, for all four misfit combinations considered, mobility of screw dislocations remains sizably lower than that of edge dislocations thus casting doubts on a popular assertion that strength of complex BCC alloys may be controlled by edge dislocations [38–40]. Focusing on screw dislocations, their contribution to SSS is observed to correlate with the density of debris defects left in the wake of moving screw dislocations (not shown): the debris density is observed to be notably greater in two stronger alloys, the solo stiffness misfit alloy (0, 0.3) and the synergistic (0.03, 0.3) B1 alloy. Observing that size and stiffness misfits have qualitatively similar effects on alloy strength and on mobility of individual dislocations, we leave for future work a more detailed quantitative assessment of the relationship between dislocation mobility and plastic strength.

Our extensive exploration of the effects of two misfits on SSS has so far relied solely on EAM alloy potentials [21] in which cross-interactions among alloy elements, e.g. between Nb and W atoms, is defined by a convenient scaling rule that made it possible for us to generate alchemical elements with arbitrary combinations of size and stiffness misfits. At the same, cross-interactions among 16 metals of the same set of EAM potentials have not been fitted to properties of any specific alloys. To verify if and how much the trends predicted in our massive MD simulations for the EAM model alloys hold for other models of interatomic interactions, we performed additional massive MD simulations with Spectral Neighbor Analysis Potential (SNAP) potentials recently developed for Nb-Mo-Ta-W alloys [41,42]. By mapping local atomic environments on a set of bi-spectrum descriptors and including in their training data results of accurate density functional theory (DFT) calculations pertaining to binary alloys, the resulting SNAP potentials are expected to be more accurate than the EAMs in describing interatomic interactions in complex Nb-Mo-Ta-W alloys [42]. To our knowledge no simple scaling rule exist for the SNAP potentials similar to the one we exploited to modify atom size and elastic stiffness of the EAM potentials. Instead, each of the four elements in this family of SNAP model alloys represents a real metal and comes with its own specific values of atomic radius and elastic stiffness. To compare EAM and SNAP predictions for alloy strength it is necessary to define a measure of strengthening applicable to both interatomic potential models. Here we define relative strengthening $\delta_f$ as:

$$\delta_f = \frac{\sigma_{\text{alloy}}}{\sum_i c_i \sigma_i} - 1 \qquad (5)$$

where $\sigma_{\text{alloy}}$ is strength computed in an MD simulation of a random alloy with the element fractions $c_i$ and the denominator in the first term on the right-hand side is a composition-dependent reference strength with respect to which SSS is measured in each SNAP model alloy. Parameter $\delta_f$ quantifies the notorious "cocktail effect", namely if and how much strength of an alloy exceeds the "rule of mixture" (or Vegard's law) strength weighted by atomic fractions of alloy components. Whereas reference rule of mixture strength depends on alloy composition for SNAP model alloys, owing to constraints we imposed on constructing our EAM model alloys, their composition-averaged strength is independent of composition and remains equal to strength of the reference EAM model (elemental Ta).

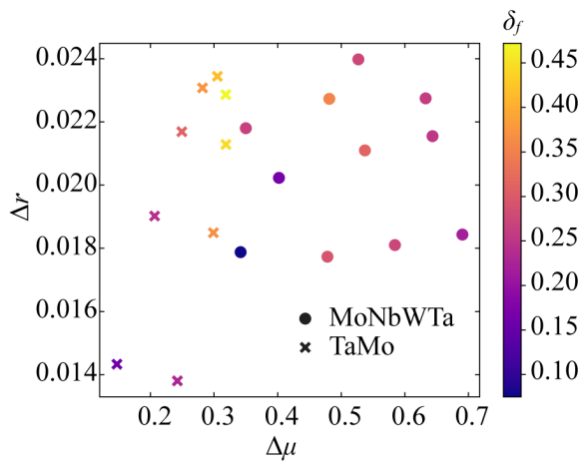

**Figure 9. Strengthening in SNAP alloys.** Relative increase in flow stress ($\delta_f$) as a function of size misfit ($\Delta r$) and stiffness misfit ($\Delta \mu$) for binary (TaMo) and quaternary (MoNbWTa) alloys modelled with the SNAP potentials.

Due to their relatively close atomic radii, the largest size misfit achievable among four elements represented in the SNAP potential (Nb, Mo, Ta, and W) is 0.025 whereas the largest achievable stiffness misfit is 0.8. For our additional MD simulations with the SNAP potentials we selected alloy compositions so as to span the plane of two misfits as widely as possible while still staying within our large yet finite computational budget. We chose to focus on Ta-Mo binary alloys (nine alloy compositions) and Nb-Mo-Ta-W quaternary alloys (11 alloy compositions). Plotted in Fig. 9 is the simulated strength of SNAP model alloys as a function of size and stiffness misfits. Similar to the EAM model alloys, relative strengthening among each SNAP alloy system increases with increasing size and stiffness misfit. Following our rule for assigning alloys to seven distinct types according to their misfit pair type combinations, all nine of simulated SNAP binary alloys and nine of 11 quaternary alloys are of the B2 type with the maximum relative strengthening among such alloys staying below 0.47. For comparison, maximum relative strengthening among EAM alloys of B2 type over the same range of sampled misfits is 0.49. An antagonistic effect of two misfits is clearly observed among 18 SNAP model alloys of the B2 type although it is less pronounced in comparison to the EAM model alloys depicted in Fig. 4b which is likely due to a considerably narrower range of variations in atom size and stiffness among four SNAP model metals. Two of 11 simulated SNAP model quaternaries fall into T1 category and do not demonstrate any prominent strengthening, again in qualitative agreement with EAM model alloys of the same type.

**Conclusion**

To carefully distinguish and quantify effects of atom size and elastic stiffness misfits on CCA strength, we devised an MD simulation approach in which potentials describing interatomic interactions are modified to create model elements with arbitrary combinations of atoms size and stiffness subsequently combined into model alloys with arbitrary size and stiffness misfits. Rather than attempting to simulate real CCAs, we rely on computational alchemy to systematically sample variations of alloy strength in response to variations in two misfits. By not trying to imitate any particular real alloy, our alchemical approach affords means and flexibility not available in the real chemical world. Subsequently, our results do not point to specific alloy chemistries and compositions of high mechanical strength, but reveal systematic trends that can inform alloy designers in which corners of the alloy design space it may be worth looking in search of mechanically strong materials. Our massive MD simulations cover a wide range of misfits limited only by crystal instability with respect to amorphization observed here under compressive deformation in alloys with large misfits. Although not studied here, it is likely that, among alchemical alloys that do remain stable against deformation-induced amorphization, some if not many would be unstable against diffusive phase transformations inhibited here by virtue of short time scales (~10 ns) of our MD simulations over which little if any atomic diffusion takes place at 300K. On the flip side, this very time limitation likely allowed us to explore an otherwise unattainable range of misfits, to bring out their effects on material strength more clearly and to better clarify the nature of SSS in concentrated alloys.

Our principal findings are:

- Within the range of two misfits studied here, stiffness misfit induces greater strengthening than size misfit. Our simulations suggest that strengthening achievable in alloys comprised solely of refractory metals is primarily caused by stiffness misfit.
- Effects of size and stiffness misfits on alloy strength can combine in several distinct ways, both synergistic and antagonistic. Our results clearly delineate misfit combinations causing such distinct behaviors.
- Our observations suggest a concise classification of concentrated alloys into seven distinct types, each type with its own strengthening response to variations in two misfits. This classification reduces complexity of alloys with arbitrary number of components to no more than four consolidated components.
- Consolidated binary alloys in which one component is small and soft and the other component is large and stiff are the most promising alloy types for enhanced mechanical strength.

- The number of elements in CCAs does not significantly affect alloy SSS, as long as the size and stiffness misfits are controlled.

Our computational approach augments the much more constrained and thus inevitably spotty experimental exploration of the CCA design space. Taking advantage of full control of model material parameters unique to our approach, our study demonstrates that useful insights can be gained from intentionally unrealistic models.

**Methods**

**1. Molecular Dynamics simulations of alloy strength and dislocation mobility**

To quantify contributions of size and stiffness misfits to alloy strengthening we devised a procedure to scale parameters of an EAM interatomic potential originally developed for Ta [21] so as to vary the element's atom size and/or elastic stiffness while leaving the ground state structure of the alchemical element the same as the base Ta potential, i.e. body-centered cubic (BCC) (see the following section for details). Atom size and elastic stiffness are defined as the equilibrium distance between the nearest neighbor atoms and the Voigt average shear modulus [43], both computed at 0K. To construct an $n$-component alloy, $n$ alchemical elements (scaled EAM potentials) are created and their atoms are distributed randomly at predefined atomic fractions $c_i$ over the sites of a BCC lattice comprised of ~55 million atoms seamlessly embedded in an infinite crystal under three-dimensional periodic boundary conditions. Oriented along the principal [100], [010], [001] axes of the BCC lattice along its $x$, $y$ and $z$ directions, each crystal had initial aspect ratio 4:1:1. All MD simulations were performed using the open-source code LAMMPS [44]. Once constructed, each model alloy was equilibrated over 100 ps at 300 K and zero pressure in an isoenthalpic-isobaric (NPH) ensemble and subjected to compressive deformation along its longest $x$-axis at a strain rate of $10^8$ s$^{-1}$ while maintaining temperature and pressure constant near 300 K and 0 bar respectively.

To capture realistic complexity of single crystal plasticity in an MD simulation it is essential to populate the crystal with initial dislocation sources [45,46]. As was amply demonstrated in our earlier work [46–48], shapes and density of initial sources affect only the transient plastic response of a crystal whereas plastic flow attained at large simulated strains does not depend on the exact configuration of initial dislocation sources. For this important reason, here we choose flow stress attained at true strains reaching 0.4 as the measure of crystal's plastic strength. When performed at such large scales and under such tightly controlled conditions, MD simulations have been shown to be statistically representative and to reproduce full details of single crystal plasticity where thousands of dislocations move and interact in their natural ways, not being prescribed how to behave but following only the underlying motion of atoms. This is in contrast to previous MD simulations of CCAs in which only dislocation-free crystals or at most a single dislocation per crystal have been considered. To compare and relate predictions of our massive MD simulations to preceding work we also ran MD simulations of single edge and screw dislocations in pure Ta and four different model alloys. Using much smaller simulation volumes containing only ~520 thousand atoms, MD simulations for individual dislocations were performed under simulation conditions described in [49] at the same temperature 300K and in the same range of applied stress typical of our massive MD simulations.

## 2. Creating model elements with variable atom size and elastic stiffness

In the EAM formalism [50], total energy $E$ of a system of interacting atoms is written as

$$E = \frac{1}{2} \sum_{i,j,i \neq j} \phi(r_{ij}) + \sum_i F(\rho_i) \tag{6}$$

where the sums are over all atoms $i$ in the model, $\phi$ accounts for the interaction energy between a pair of atoms $i$ and $j$ separated by distance $r_{ij}$, and $F$ defines the energy it takes to embed atom $i$ into electron density $\rho_i$. The latter density is computed as a sum over partial electron densities $f$ supplied at the position of atom $i$ by its neighbor atoms $\rho_i = \sum_j f(r_{ij})$.

Various functional and tabulated forms have been used for functions $\phi, F$ and $f$ since the EAM formalism was first proposed. Zhou et al. [21] fitted parameters of their three functions to basic material properties of 16 metallic elements, one parameter set for each metal. In principle, functions $\phi, F$ and $f$ describing cross-species interactions, e.g. between Ta and W atoms, could be fitted to properties pertaining to appropriate alloys. To avoid having to fit additional $16 * \frac{15}{2} = 120$ cross-species potentials among 16 elements, the authors kept functions $F_a, \phi_{aa}$ and $f_a$ for each metal $a$ unchanged, but expressed pair interactions among species $a$ and $b$ as follows

$$\phi_{ab}(r) = \frac{1}{2} \left[ \frac{f_b(r)}{f_a(r)} \phi_{aa}(r) + \frac{f_a(r)}{f_b(r)} \phi_{bb}(r) \right] \tag{7}$$

Although possibly not entirely realistic, this form of cross-interactions enables atomistic simulations of alloys comprised of arbitrary combinations of 16 metals.

Essential for our purposes here was that we could use the above averaging rule to define cross-interactions among our model alchemical elements. In this study, we relied on Zhou et al. [21] parameterization of their EAM potential for Ta as a reference element and then varied its EAM parameters so as to modify atomic size and/or stiffness. Specifically, to create a model element of atomic size $x$ times larger and/or $y$ times stiffer than Ta, we scaled the equilibrium distance $r_e$ parameter of the Zhou at al. EAM potential for Ta by a factor of $x$ and/or scaled all 11 parameters of the same potential with the dimension of energy by a factor of $x^3 y$. This simple approach allowed us to generate model elements with tailored size and stiffness and to assemble model alchemical alloys with specific size and stiffness misfits.

## Acknowledgments

The authors acknowledge useful discussions with J. McKeown, A. Perron, T. Voisin, J. Marian, P. S. Branicio, and N. R. Barton. The authors acknowledge funding support from the ASC PEM program at Lawrence Livermore National Laboratory. Computing support for this work came from LLNL Institutional Computing Grand Challenge program. This work was performed under the auspices of the U.S. Department of Energy by Lawrence Livermore National Laboratory under Contract DE-AC52-07NA27344.

## Author contributions

Conceptualization: A.L. and V.V.B.; Methodology: A.L., N.B, X.Z., S.A., and V.V.B.; Investigation: A.L., N.B.; Visualization: A.L., N.B.; Writing – Original Draft: A.L., N.B., and V.V.B.; Writing – Review & Editing: A.L., N.B., X.Z., S.A., and V.V.B; Supervision: V.V.B.; Funding Acquisition: V.V.B.

## Declaration of interests

The authors declare no competing interests.